# The features of contact resistivity behavior at helium temperatures for InP- and GaAs-based ohmic contacts

A.V. Sachenko[1], A.E. Belyaev[1], N.S. Boltovets[2], S.A. Vitusevich[3, a)],

R.V. Konakova[1], S.V. Novitskii[1], V.N. Sheremet[1]

[1]V. Lashkaryov Institute of Semiconductor Physics, NAS of Ukraine, 41, Nauki prospect, 03680, Kyiv, Ukraine

[2]State Enterprise Research Institute "Orion", 8a Eugene Pottier St., 03057, Kyiv, Ukraine

[3] Peter Grünberg Institute, Forschungszentrum Jülich, 52425 Jülich, Germany

Contact resistivity $\rho_c$ of InP- and GaAs-based ohmic contacts was measured in the 4.2–300 K temperature range. Nonmonotonic dependences $\rho_c(T)$, with a minimum at $T = 50$ K (150 K) for InP (GaAs)-based contacts were obtained. The results can be explained within the framework of the mechanism of current flow through metal shunts (associated with dislocations) penetrating into the semiconductor bulk, with allowance being made for electron freeze-out at helium temperatures. Contact ohmicity in the 4.2–30 K temperature range is due to accumulation band bending near shunt ends at the metal−semiconductor interface.

## I. INTRODUCTION

As yet only a few papers have been published dealing with the measurement of contact resistivity $\rho_c$ for GaAs-based metal−semiconductor ohmic contacts in a wide (from helium to room) temperature range (Ref. 1,2). We are not aware of any similar work on InP-based contacts. At the same time, the manufacturing technologies for ohmic contacts (operating at helium temperatures) based on III–V compounds require meticulous development. It is well known that a number of semiconductor devices which ensure functioning of superconductive systems operate at very low temperatures (Ref. 3). Until recently, the only known mechanism of current flow capable of ensuring contact ohmicity at such temperatures was the purely

---

a) Electronic mail: s.vitusevich@fz-juelich.de. On leave from the Institute of Semiconductor Physics, NASU, 03028 Kiev, Ukraine.

field mechanism. However, the realization of this mechanism requires the application of either rather heavily doped semiconductors degenerate at helium temperatures or a doping step. At the same time, it was shown in (Ref. 1,2) that contact ohmicity is retained at helium temperatures for nondegenerate semiconductors as well. In that case, the thermionic or thermal field mechanism of current flow cannot ensure contact ohmicity in the 4–10 K temperature range.

In (Ref. 4-7), another mechanism was proposed that ensures contact ohmicity at any temperatures (including helium temperatures). The principle of this mechanism is based on current flowing through metal shunts that penetrate deep into the semiconductor bulk. In that case, the necessary condition for contact ohmicity is realization of accumulation band bending for majority charge carriers in the semiconductor region adjacent to the metal shunt. Then there is no additional drop in contact voltage in the near-contact space charge region (SCR) of the semiconductor, i.e., the contact is in fact ohmic. In that case, however, the flowing current is limited by the diffusion supply of electrons. As a result, the contact resistivity $\rho_c$ is inversely proportional to the electron mobility $\mu_n$. This, in particular, ensures the growth of $\rho_c$ with temperature $T$.

It was shown in (Ref. 4) that the presence of accumulation band bending in the semiconductor at the boundaries with metal shunt ends can be ensured by the mirror image forces on condition that the geometrical sizes of the shunts are close to the atomic sizes. Such a situation may be realized, in particular, for the dislocations through which the metal shunts grow. It was stressed by Gol'dberg et al. (see, e.g., the review (Ref. 8)) that the shunts can penetrate into semiconductor bulk via other extensive defects than dislocations, in particular, via stacking faults and intercrystalline boundaries. It should be noted that we performed a model experiment (Ref. 5) to prove that the current flows through metal shunts associated with dislocations (at a rather high dislocation density) lead to contact ohmicity. Firstly, the silicon samples with different doping levels were lapped. Then palladium was deposited to form contact pads. As a result of such material treatment the obtained contacts were ohmic, while the Pd-Si contacts formed without lapping demonstrated Schottky-type behavior (Ref. 9). The dislocation density was measured to be of $10^7$ cm$^{-2}$ for a lapped silicon surface. It should be emphasized that the value of conducting dislocation density required for agreement between the theory and experiment was found to be the same order of magnitude.

In the present work, we studied the temperature dependence of contact resistivity $\rho_c$ for GaAs- and InP-based ohmic contacts in the 4.2–300 K temperature range. It is shown that, with assumption made for electron freeze-out at very low temperatures, the experimental $\rho_c(T)$ curves are well described using the current flow mechanism proposed in (Ref. 4-7).

## II. EXPERIMENTAL PROCEDURE

The objects of investigation were the Au (200 nm)–TiB$_2$ (100 nm)–Au (180 nm)–Ge (30 nm)–$n$-$n^+$-$n^{++}$-GaAs and Au (200 nm)–TiB$_2$ (100 nm)–Ge (40 nm)–Au (180 nm)–$n$-$n^+$-$n^{++}$-InP ohmic contacts. They were produced using magnetron sputtering of metals and TiB$_2$ onto the GaAs- and InP-based $n$-$n^+$-structures heated up to 100 ºC. The latter were VPE-grown on the $n^{++}$-GaAs(100) and $n^{++}$-InP(100) substrates. The impurity concentration in the heavily doped $n^{++}$-GaAs(100) substrates was $2 \times 10^{18}$ cm$^{-3}$, while in the $n^{++}$-InP(100) substrates it was about $10^{18}$ cm$^{-3}$. The thickness of both types of substrates was 300 μm. The impurity concentration in the 3 μm thick buffer $n^+$-layer of GaAs (InP) was $5 \times 10^{17}$ cm$^{-3}$, while that in the 3 μm thick $n$-layer of GaAs was $6 \times 10^{15}$ cm$^{-3}$ and $9 \times 10^{15}$ cm$^{-3}$ in the 2 μm thick $n$-layer of InP. The ohmic contacts to GaAs (InP) were formed using rapid thermal annealing at $T = 440$ °C for 60 s ($T = 460$ °C for 30 s).

Contact resistivity $\rho_c$ of the above ohmic contacts in the packaged samples was measured in the 4.2–300 K temperature range by the transmission line method at voltages don't exceeding value of 100 mV.

## III. DISCUSSION OF RESULTS

Figures 1 and 2 present experimental $\rho_c(T)$ curves for InP- and GaAs-based ohmic contacts. It can be seen that the $\rho_c(T)$ curves have a minimum (at $T$ of about 50 K for InP and 150 K for GaAs). The dependences $\rho_c(T)$ grow on both sides of the minimum (more rapidly on the left side, as the temperature decreases).

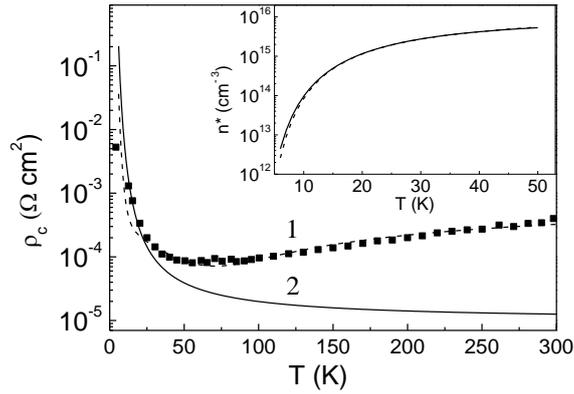

Fig. 1. Temperature dependences of contact resistivity for InP-based ohmic contacts (impurity concentration of $9\times10^{15}$ cm$^{-3}$): black squares - experiment, solid curve - calculation (with parameters $N_{D1} = 1.7\times10^8$ cm$^{-2}$, $N_{D2} = 10^4$ cm$^{-2}$). Inset: temperature dependence $n^*(T)$ for InP-based ohmic contact (activation energy of 4.5 meV).

Neither the thermionic nor the thermal field mechanism of current flow can explain the experimental dependences $\rho_c(T)$ or ensure the ohmicity of contacts based on III–V compounds at helium temperatures. In fact, the realization of the thermionic mechanism of current flow per se is not at variance with contact ohmicity. However, the $\rho_c(T)$ curve grows rapidly at temperatures ≤50 K, which is characteristic of Schottky contacts.

Really, despite low doping level of semiconductors in the structures studied by us, the contact resistance $R_c = \rho_c/S$ (where $\rho_c$ is the specific contact resistivity and $S$ is the contact area) and bulk resistance $R_b$ were of the same order of magnitude over the whole range of temperatures, including helium ones. This can be seen from the following data (insert to Fig. 2) presenting $R_c/R_b$ ratio as function of temperature for the GaAs-based contacts (black squares) and InP-based contacts (open circles).

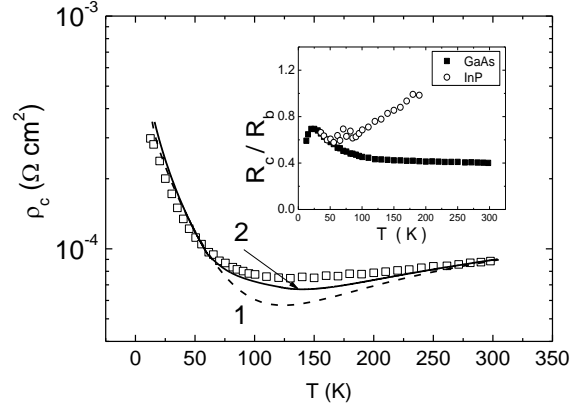

Fig. 2. Temperature dependences of contact resistivity for GaAs-based ohmic contacts (impurity concentration of $6\times10^{15}$ cm$^{-3}$): open squares - experiment, curves - calculation (with parameters $N_{D1} = 3.7\times10^{7}$ cm$^{-2}$, $N_{D2} = 10^{4}$ cm$^{-2}$). Insert: experimental dependencies $R_c/R_b(T)$: black squares - GaAs, open circles – InP-based contacts.

In that case, presence of Schottky barrier contacts will lead to non-ohmicity of the I-V curves, i.e., the contacts would be non-ohm. Besides, if the $\rho_c(T)$ curves are growing (such is our case), this indicates contact ohmicity independent of the ratio between $R_c$ and $R_b$.

To prove the above statement, let us use the expression for $\rho_c(T)$ that allows for thermal tunneling (Ref. 10):

$$\rho_c(T) = C \exp \frac{\varphi_b}{E_{00} \coth(E_{00}/kT)}. \qquad (1)$$

Here $E_{00} = 0.054\left((m_0/m^*)(n_0/10^{20}\text{ cm}^{-3})(11.7/\varepsilon_s)\right)^{0.5}$ eV is the characteristic tunneling energy, $C$ a constant, $\varphi_b$ barrier height, $m^*$ and $m_0$ electron effective mass and free electron mass, respectively, and $\varepsilon_s$ semiconductor permittivity.

Fig. 3 shows the calculated $\rho_c(T)$ curve in the 5−300 K temperature range constructed by using Eq. (1) for an InP-based contact (impurity concentration of $9\times10^{15}$ cm$^{-3}$), as well as the experimental dependence $\rho_c(T)$. The theoretical curve 1 is constructed using the barrier height value $\varphi_b = 10$ meV, on the assumption that $n_0$ does not depend on temperature. By choosing an appropriate $C$ value, it is possible to bring the calculated dependence $\rho_c(T)$ into agreement with the experimental dependence in a narrow (5−40 K) temperature range. At higher temperatures, the theoretical and experimental dependences diverge drastically. For example, the experiment yields a growth of $\rho_c$ with temperature, while the theoretical $\rho_c(T)$ curve decays. However, even agreement between theory and experiment at low temperatures is only apparent. Eq. (1) at $T < 50$ K cannot be applied because it does not take electron freeze-out at low temperatures into account. In this case, it seems more correct to use $n^*(T)$ (rather than $n_0$), which is determined from the following neutrality equation:

$$n^*(T) = \frac{n_0}{1 + 2\exp((E_f - E_d)/kT)} = \frac{2}{\sqrt{\pi}} N_c \left(\frac{T}{300\,\text{K}}\right)^{3/2} \int_0^\infty \frac{x^{0.5}}{1 + \exp(x - E_f/kT)} dx. \qquad (2)$$

Here $E_f$ is the Fermi energy, $E_d$ donor energy level, and $N_c$ effective density of states in the conduction band at $T = 300$ K. However, it can be seen from Fig. 3 that, with assumption being made for electron freeze-out, there is no agreement between theory and experiment even at low temperatures (see curve 2).

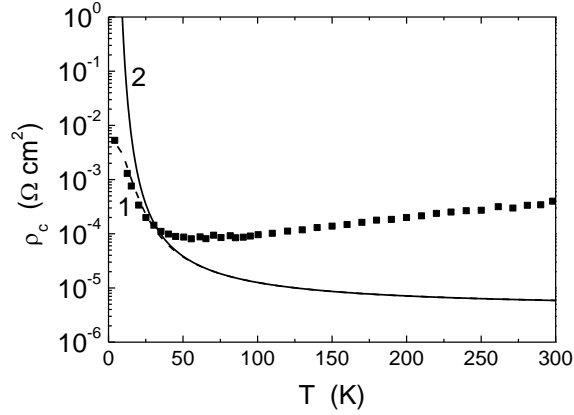

Fig. 3. Dependences $\rho_c(T)$ calculated using Eq. (1) (solid curves) and experimental dependence $\rho_c(T)$ for InP-based contacts (marked). Curve 1 (2) is constructed without (with) allowance being made for electron freeze-out; the barrier height $\varphi_b = 10$ meV.

Application of the approach developed in (Ref. 4-8), with assumption being made for electron freeze-out at low temperatures, makes it possible to describe the experimental dependences $\rho_c(T)$ obtained for InP- and GaAs-based contacts (see Figs. 1 and 2). First, we obtain an expression for $\rho_c(T)$ in the case of current flowing through metal shunts associated with dislocations. (We restrict ourselves to considering nondegenerate semiconductors.) The area from which the thermionic current flowing through a single shunt is collected equals $\pi L_D^2$, where

$$L_D = \left[\varepsilon_0 \varepsilon_s kT / 2q^2 n^*(T)\right]^{1/2} \qquad (3)$$

is the Debye screening length, $\varepsilon_0$ vacuum permittivity, and $k$ Boltzmann constant.

Let us consider that accumulation band bending is realized at a dislocation outcrop in the semiconductor. The density $J_{nc}$ of thermionic current flowing through the contact at the dislocation outcrop can be determined by solving the continuity equation for electron current. The interrelation between the bulk electron concentration, $n^*(T)$, and nonequilibrium electron concentration $n(x)$ at a point $x$ of the near-contact SCR is determined by double integration of the continuity equation for electron current over the coordinate $x$ perpendicular to the metal−semiconductor interface. In the case of a nondegenerate semiconductor,

$$n(x) = e^{y(x)} \left( n^*(T) - \frac{J_{nc}}{qD_n} \int_x^w e^{-y(x')} dx' \right), \quad (4)$$

where $y(x) = q\varphi(x)/kT$ is the nonequilibrium nondimensional potential at a point $x$, $D_n$ is the electron diffusion coefficient, and $w$ width of near-contact SCR.

$J_{nc}$ is determined as

$$J_{nc} = q\frac{V_T}{4}(n_c - n_{c0}), \quad (5)$$

where $V_T$ is the mean electron thermal velocity, $n_c$ ($n_{c0} = n^*(T)\exp y_{c0}$) the nonequilibrium (equilibrium) electron concentration at the contact plane, and $y_{c0} = q\varphi_{c0}/kT$ equilibrium nondimensional potential at the metal−semiconductor interface.

Setting $x$ in Eq. (4) equal to zero and using Eq. (5), we determine $n_c$. Substituting the obtained expression for $n_c$ into Eq. (5) and taking into account that the nonequilibrium nondimensional potential $y_c = y_{c0} + \ln(qV/kT)$ (this is the condition of contact ohmicity), we obtain the final expression for the density of electron current flowing through the metal−semiconductor contact at the dislocation outcrop:

$$J_{nc} = V/\rho_{c0}, \quad (6)$$

where

$$\rho_{c0} = \frac{kT}{q} \frac{\left(1 + \frac{V_T}{4D_n} e^{y_{c0}} \int_0^w e^{-y(x)} dx\right)}{\frac{qV_T}{4} n^*(T) e^{y_{c0}}}. \quad (7)$$

The integral $\int_0^w e^{-y} dx$ may be presented in a form that is more convenient for calculations when passing from integration over coordinate $x$ to integration over nondimensional potential $y$:

$$\int_0^w e^{-y}dx = L_D \int_{y_c}^{y_x} \frac{e^{-y}dy}{\left(e^y - y - 1\right)^{0.5}}. \tag{8}$$

The calculation shows that the value of the integral in Eq. (8) (in the case of $y_x = 0.5$) varies from 0.56 (at $y_{c0}$ = 1.5) to 0.65 (at $y_{c0}$ = 3.5) and practically flattens out at higher $y_{c0}$ values,

The contact resistance determined by the mechanism of diffusion supply of electrons can be found (for a contact of unit area) from the following expression:

$$\rho_c(T) = \rho_{c0} / \pi L_D^2 N_{D1}, \tag{9}$$

where $N_{D1}$ is the density of conducting dislocations (those with which metal shunts are associated). Generally, the density $N_{D1}$ of conducting dislocations and that of scattering (i.e., without associated metal shunts) dislocations ($N_{D2}$) are not equal to each other. The conducting dislocations are predominantly those perpendicular to the contact–semiconductor interface, while the scattering dislocations are predominantly those parallel to that interface. The quantity $\pi L_D^2 N_{D1} S$ is the total area from which the total current flowing through all metal shunts associated with conducting dislocations is collected. As a rule, $\pi L_D^2 N_{D1}$ is much less than unity, even at maximal dislocation densities (about $10^{10}$–$10^{11}$ cm$^{-2}$), except in the case of low-doped semiconductors with $N_d \leq 10^{15}$ cm$^{-3}$.

According to Einstein's relation, the electron diffusion coefficient is $D_n = kT\mu_n / q$. Here the electron mobility $\mu_n$ is determined with allowance being made for the three main mechanisms of electron scattering: on charged impurities ($\mu_Z$), optical lattice vibrations ($\mu_o$) and dislocations ($\mu_D$):

$$\mu_n = \left(\mu_Z^{-1} + \mu_o^{-1} + \mu_D^{-1}\right)^{-1}. \tag{10}$$

In our calculations, for $\mu_Z$ and $\mu_o$ we used the expressions given in (Ref. 11) and for $\mu_D$ that given in (Ref. 12). The expressions for $\mu_Z$, $\mu_o$ and $\mu_D$ are therefore as follows:

$$\mu_z(T) = \frac{3.68 \times 10^{20} \left(\frac{\varepsilon_s}{16}\right)^2 \left(\frac{T}{100\text{K}}\right)^{3/2}}{n^* \left(\frac{m^*}{m_0}\right)^{1/2} \log\left\{1 + \left[\left(\frac{\varepsilon_s}{16}\right)\left(\frac{T}{100\text{K}}\right)\left(\frac{2.35 \times 10^{19}}{n_w \text{cm}^{-3}}\right)^{1/3}\right]^2\right\}}, \tag{11}$$

$$\mu_o(T) = \frac{31.8 \sinh\left(\frac{\theta}{2T}\right)}{\left(\frac{1}{\varepsilon_{sh}} - \frac{1}{\varepsilon_{sl}}\right)(\theta)^{1/2} \left(\frac{m^*}{m_0}\right)^{1.5} \left(\frac{\theta}{2T}\right)^{1/2} K_1\left(\frac{\theta}{2T}\right)}, \tag{12}$$

where $n_w$ is the ionized impurity concentration, $\theta$ longitudinal optical phonon temperature, $\varepsilon_{sh}$ ($\varepsilon_{sl}$) semiconductor high (low)-frequency permittivity, $K_1(\theta/2T)$ modified Bessel function of the first order, and

$$\mu_D = \frac{B \exp(\eta)}{T^{1/2} N_{D2} L_D^5} K_2(\eta), \qquad (13)$$

where $\eta = \dfrac{\hbar^2}{16m L_D^2 kT}$, and $K_2(\eta)$ is the modified Bessel function of the second order. A dimensional coefficient $B$ is

$$B = \frac{\left(\hbar^2 \varepsilon_0 \varepsilon_{sl} c\right)^2}{8\sqrt{2\pi k}\, q^3 \sigma^2 m^{*5/2}},$$

where $\sigma = \lambda/2qc$, $\lambda$ is the linear charge density of the dislocation line, and $c$ lattice constant along the [0001] direction.

It should be noted that the mobilities in Eqs. (11)–(13) are measured in cm$^2$/V×s.

The quantity

$$\beta = \frac{V_T}{4 D_n} e^{y_{c0}} \int_0^w e^{-y(x)} dx \qquad (14)$$

determines the degree of diffusion limitation that is essential at $\beta>1$.

Eqs. (2)–(13) make it possible to calculate $\rho_c(T)$ in the case of current flowing predominantly through the metal shunts associated with conducting dislocations and limited by the diffusion supply of electrons.

A distinction between the above relations and those obtained in (Ref. 4) is that the bulk electron concentration $n^*(T)$ is not considered to be constant: it decreases at low temperatures because of electron freeze-out. The inset to Fig. 1 shows the dependences $n^*(T)$ for InP-based contacts obtained in the 4–50 K temperature range. It can be seen that, at a typical $E_d$ value of 7 meV, $n^*(T)$ decreases by about an order and a half in the above temperature range ($n_0 = 9 \times 10^{15}$ cm$^{-3}$). In the above cases, the theoretical dependence $n^*(T)$ is well approximated by the activation energy $E_a$ of 4.5 meV.

Let us use the above theoretical relations to calculate $\rho_c(T)$ for a contact based on high-resistance InP ($n_0 = 9 \times 10^{15}$ cm$^{-3}$). In this case, the use of values $N_{D1} = 1.7 \times 10^8$ cm$^{-2}$ and $N_{D2} = 10^4$ cm$^{-2}$ in the calculations gives a good quantitative agreement between the theoretical (curve 1) and experimental dependences.

The dislocation density within near-contact region of InP has been measured by SEM. It was found to be of order of $10^8$ см$^{-2}$. This value is in good agreement with estimated value obtained from fitting of experimental data by theoretically calculated. This fact of the coincidence for these values of dislocation density additionally supports the realization of the dislocation mechanism for ohmic contact formation.

At low temperatures ($T$ about or less than 50 K), $\rho_c(T)$ decreases exponentially with temperature. This is rather well approximated by the following expression:

$$\rho_c(T) = A\exp(\varphi_b / kT), \qquad (15)$$

where the effective barrier height $\varphi_b$ for an InP-based contact is 6 meV (see curve 2). In the cases considered, the presence of an exponential portion in the $\rho_c(T)$ curve is not related to the realization of the thermionic mechanism of contact resistivity formation, but is determined generally by the effect of electron freeze-out at helium temperatures. The main reason for the growth of contact resistance with decreasing temperature is the presence of the portion of exponential reduction of the electron concentration in the $n^*(T)$ curve, as shown above.

We now discuss the results of the comparison between the above theory and the experimental dependences $\rho_c(T)$ for GaAs-based contacts. In that case, use of the bulk electron concentration value of $6\times10^{15}$ cm$^{-3}$ enabled us to obtain only qualitative (rather than quantitative) agreement with the experiment, in particular, nonmonotonic dependences $\rho_c(T)$. In our opinion, the reason for this is as follows. The structure of the contact under investigation is Au (200 nm)–TiB$_2$ (100 nm)–Au (180 nm)–Ge (30 nm)–$n$-$n^+$-$n^{++}$-GaAs. In this case, the 30 nm Ge layer is in direct contact with the high-resistance bulk. Since Ge serves as the donor in GaAs, the electron concentration in the near-contact layer increases as the ohmic contact is formed at $T = 440$ °C. This was taken into account and $n_0 = 2\times10^{16}$ cm$^{-3}$ was used when constructing the theoretical dependence $\rho_c(T)$. As a result, we obtained sufficient agreement between the theory (curve 1) and experiment by using $N_{D1} = 3.7\times10^7$ cm$^{-2}$ and $N_{D2} = 10^4$ cm$^{-2}$. Better agreement is achieved if the electron concentration gradient appearing due to Ge diffusion into the semiconductor is taken into account - see curve 2. This was obtained by using $n_0$ values of $5\times10^{16}$ cm$^{-3}$ at $T \leq 80$ K and $2.9\times10^{16}$ cm$^{-3}$ at $T > 100$ K.

The electrons bringing about current flow are collected at distances of about the Debye screening length $L_D$. Since this is proportional to $\sqrt{T}$, its value at helium temperatures is several times lower than at room temperature. Therefore, at helium temperatures, the electron concentration $n_0$ is higher. Fitting by applying different $n_0$ values therefore yields better agreement with the experiment.

The results of agreement between the theoretical and experimental dependences $\rho_c(T)$ in the contacts under investigation show that the density of scattering dislocations $N_{D2}$ is much lower than that of conducting dislocations. One of the possible explanations is that almost complete relaxation of misfit dislocations occurs at the contact firing temperatures used. Another explanation involves participation of other extensive defects (e.g., intercrystalline boundaries with metal shunts grown into a semiconductor) in current flow.

## IV. CONCLUSION

Nonmonotonic dependences $\rho_c(T)$ were obtained for rather high-resistance GaAs- and InP-based ohmic contacts in the 4.2–300 K temperature range. The results of our work show that the experimental dependences $\rho_c(T)$ are well described by theoretical curves calculated under assumption of current limitation by diffusion supply of electrons, with allowance made for low-temperature freeze-out of electrons. It should be emphasized that the fundamental mechanism is based on the realization of accumulation band bending at the ends of semiconductor space charge regions bordering with metal shunts. This effect allows explaining the contact ohmicity. The attempts to relate the results obtained to thermionic mechanism of current flow in weakly-rectifying contacts are untenable, in spite of presence of exponential portion of the $\rho_c(T)$ curve. The latter appears because of reduction of concentration of mobile electrons due to low-temperature freeze-out of electrons.